\documentclass[twocolumn,floatfix,amsmath,amssymb,aps,prb,showpacs]{revtex4}
\usepackage{epsfig}
\usepackage{amsmath}

\usepackage{graphicx}
\usepackage{color}

\usepackage[normalem,normalbf]{ulem}
\usepackage{cancel}

\begin{document}

\title{Phase diagram of Model C in the parametric  space of order parameter and space dimensions}

\author{ M. Dudka$^1$, R. Folk$^2$, Yu. Holovatch$^1$, }
\affiliation{
$^1$ Institute for Condensed Matter Physics, National
Academy of Sciences of Ukraine, Svientsitskii str. 1, UA--79011 Lviv, Ukraine\\
$^2$ Institut f\"ur Theoretische Physik, Johannes Kepler
Universit\"at Linz, A--4040, Linz, Austria
}
\date{\today }

\begin{abstract}
The  scaling behavior of  model C  describing the dynamical behaviour of the $n$-component
nonconserved order parameter coupled statically to a scalar conserved density  is considered in
$d$-dimensional space.  Conditions for the realization of different types of scaling regimes
in the $(n,d)$ plane  are studied within the field-theoretical renormalization
group approach. Borders separating these regions  are calculated  on the base of high-order RG functions
using $\epsilon$-expansions as well as by fixed dimension $d$ approach with resummation.


\end{abstract}

\maketitle

\section{Introduction}
According to the dynamical universality hypothesis the dynamical properties  of physical systems in the vicinity of their critical points
can be grouped into universality classes (similar to static ones)  irrespective of details of their microscopic dynamical behavior.
There, in addition to the global parameters of a system, crucial role is played by the behavior of the relevant slow variables,
namely order parameter and secondary densities associated with conservation laws. Their dynamical behavior is described by a set of equations of
motion of Langevin type with Gaussian noise terms caused by remaining microscopic degrees of freedom \cite{Halperin77,Folk06,Tauber}.

The coupling between the order parameter and secondary densities is also important.
In the simplest case  such coupling is realised within the so called model C \cite{Halperin74,Halperin77},
where a nonconserved order parameter is coupled via a static term only to a scalar conserved density. Being quite simple, the
model can be used to describe different physical systems.
In particular,   a  lattice model of intermetallic alloys \cite{Corentsveig97},  the supercooled liquids  \cite{Tanaka99},
layers on solid substrates \cite{Binder82} are described by models with nonconserved order parameter and additional coupled conserved density. Systems
containing annealed impurities with long relaxational
times \cite{Grinstein77} manifest certain similarity with the model C
as well. {  It was argued that relativistic scalar field theory in 2+1 dimensions is consistent with model C dynamics \cite{Berges10}}.
Recently model C was applied for description of the solid-liquid like phase transition \cite{Castillo15}.
Numerical simulations of model C critical dynamics were  performed for an
Ising antiferromagnet with conserved  full magnetization and
non-conserved staggered magnetization  \cite{Sen98,Zheng01} and also for an Ising magnet with conserved
energy \cite{Stauffer}.

In order to describe critical properties of the system it is standard now  to apply  renormalization group (RG) methods.
The dynamical universality class of model C was studied  first within the dynamical version of perturbative field-theoretical
RG \cite{Folk06,Tauber}.
The behavior of model C  with an $n$-component order parameter was analyzed by $\epsilon=4-d$ expansion in different regions of  $(n,d)$ plane in the first order\cite{Halperin74} and subsequently in two loop order \cite{Brezin75,Murata76}.
Results of Ref.~\cite{Brezin75} were corrected by later calculations \cite{Folk03}.

\begin{figure}[!htbp]
\includegraphics{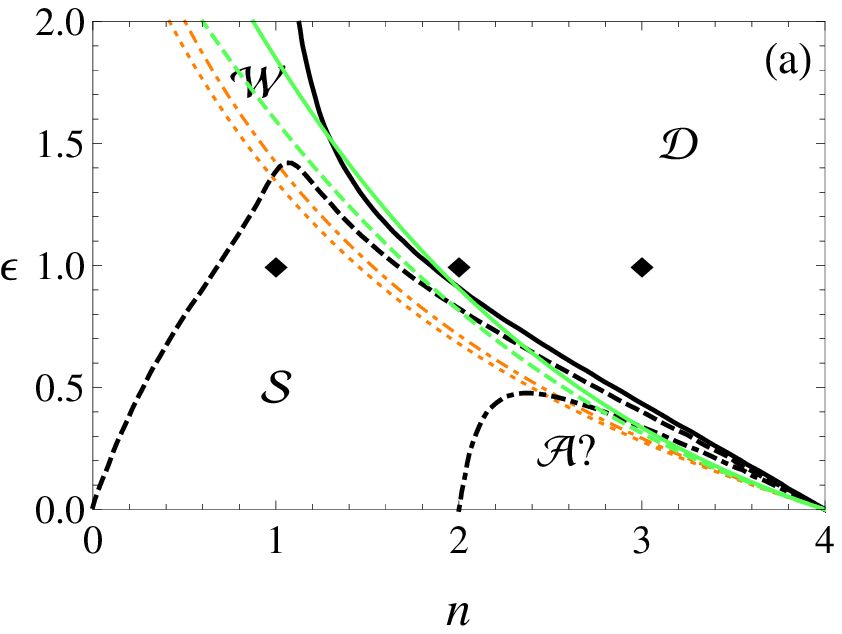}
\includegraphics{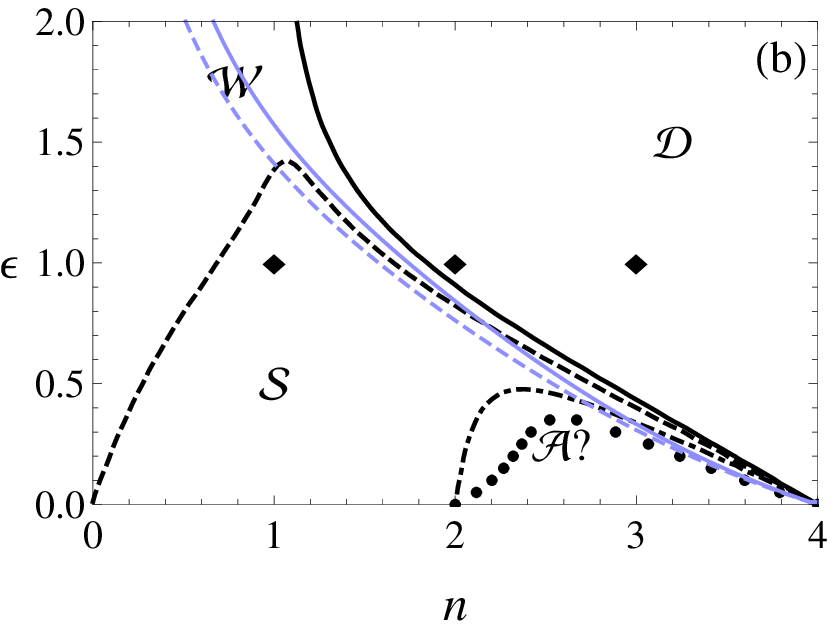}
\caption{
Phase diagram for model C in the plane ($n,\epsilon=4-d$) obtained within the perturbative field-theoretical approach  by resummation of the $\epsilon$-expansion ({\bf a}) and due to resummation of the RG functions at fixed $d$ ({\bf b})  in comparison with the non-perturbative results \protect\cite{Mesterhazy13}. Solid lines separate the decoupled region ${\mathcal D}$ from the weak scaling region $ \mathcal W$,  dashed lines separate the weak scaling region $ \mathcal W$ from the strong scaling region $ \mathcal S$.  Green curves ({\bf a}):  Pad\'e-Borel resummation of
epsilon-expansions, data of this paper; blue curves ({\bf b}):  fixed dimension approach based on resummed five-loop RG functions, data of this paper; black curves: nonperturbative RG analysis, data from Ref.~\protect\cite{Mesterhazy13}. Two loop order results of Ref. \protect\cite{Folk03} are presented as well in ({\bf a}) by dot dashed and dotted orange curves.    The anomalous region $\mathcal A$ predicted by  the nonperturbative RG is shown separated by the  black dot-dashed line (data from Ref. \protect\cite{Mesterhazy13}),  for the meaning of the black dots see the text. The diamonds mark locations of three-dimensional Ising, XY, and Heisenberg models in  ($n,d$) plane.
}
\label{curves}
\end{figure}

According to the two-loop  results \cite{Folk03} three different regimes are observed within $(n,\epsilon=4-d)$ plane (see Fig.~\ref{curves}): (1) {\em decoupled regime} (region ${\mathcal D}$),
where the secondary density decouples from the order parameter  and therefore the order
parameter dynamics  is appropriately described by model A with the dynamical critical exponent $z=2+c \eta$, where $\eta$ is the critical exponent of the pair correlation function and $c$ is some coefficient,
 while the dynamical exponent for the secondary density is $z_m=2$; (2)
{\em weak scaling regime} (region ${\mathcal W}$), where the  order parameter and the  secondary density scale differently,
the order parameter with  $z=2+c \eta$,
and the secondary density with $z_m=2+\alpha/\nu$,
where $\alpha$ and $\nu$ are critical exponents of the specific heat and of the correlation length correspondingly; {(3)} {\em strong scaling regime} (region ${\mathcal S}$),
where the order parameter and  the secondary density scale both with the  critical exponent $z=z_m=2+\alpha/\nu$ of the conserved density. {  The border lines obtained within two loop approximations \cite{Folk03} are shown in Fig~\ref{curves}a by orange curves: the dot dashed curve means border between ${\mathcal D}$ and ${\mathcal W}$, while the dotted one means
border between ${\mathcal W}$ and ${\mathcal S}$.}

There is a question about the existence of one more region (4) of the so-called {\em anomalous  scaling regime} (region ${\mathcal A}$) discovered within one-loop order \cite{Halperin74}
for $d$ near 4  for $2<n<4$. In this region the order parameter  behaves much faster then the secondary density
and scaling is questionable.  The region ${\mathcal A}$ was shown to be an artifact of the $\epsilon$-expansion within  correct two-loop calculations \cite{Folk03}.

An alternative  approach for investigating critical dynamics is the nonperturbative RG (NPRG).
This approach is based on the use of the exact RG equation for an effective action \cite{Berges02,NPRG}. It is developed intensively now and  is  applied to
describe scaling properties of different classical and quantum models.
In particular it is  successful in studies of the critical properties of  $O(n)$ models \cite{Berges02,Jakubczyk14,Codello13,Mati15},
models with different types of disorder \cite{disorder}, systems with complex symmetries \cite{Delamotte04,Delamotte15,multy},
critical dynamics near equilibrium \cite{Canet}, reaction-diffusion processes \cite{reaction}, and fully developed turbulence \cite{turbulence} to mention some examples.

Recently the NPRG approach  was also applied to study model C dynamics  \cite{Mesterhazy13}.
These results show changes concerning the  borders separating  different regimes in the $(n,d)$ plane {(shown by black curves in Fig.~\ref{curves})} as well as they report the existence of a new region ${\mathcal A}$ {(shown by dot dashed black curve in Fig.~\ref{curves})}.

In this paper, to elucidate the discrepancy between both RG approaches  we reconsider model C within the perturbative RG approach. Special attention is paid to the border lines separating different scaling regimes  in the $(n,d)$ plane. We use results of  high orders of
perturbation theory completed  by resummation procedures. { Our results show the qualitative shift up in values of $\epsilon$ for
fixed $n$ as well as the larger region $\mathcal W$  (see green curves in Fig.~\ref{curves}a  and blue curves in Fig.~\ref{curves}b) in comparison with two-loop order calculations of Ref.~\cite{Folk03} (see orange curves in Fig.~\ref{curves}a ). Therewith this outcome qualitatively supports these results of the NPRG study \cite{Mesterhazy13} (see black curves in Fig.~\ref{curves}) obtained  for $n>1$ and $\epsilon<1$,  showing them trustable. However our results do not confirm the existence of region $\mathcal A$  found in NPRG calculations  \cite{Mesterhazy13} (see black dot-dashed curve in Fig.~\ref{curves}). There are also severe differences with NPRG results in a rest of $(n,\epsilon=4-d)$ plane. In particular, the non-monotonic behavior of the border line between regions $\mathcal W$ and $\mathcal S$ (see black dashed curve in Fig.~\ref{curves}) for $n<1$ is not observed within our approach. That indicates that  NPRG treatments require improvements also concerning model A.}

 The rest of the paper is
organized as follows: in Section \ref{II} we present
dynamical model C and its RG description. Then
we analyze the conditions for realization of different
regimes of critical dynamics. We present results obtained on the base of high-order $\epsilon$-expansion
 in Section \ref{III}. We devote
Section \ref{IV} to results obtained by resummation of static five-loop RG functions of minimal subtraction scheme.  We
end the paper with a discussion and a conclusion in Section \ref{V}.

\section{Model and its RG description \label{II}}
Model C introduced  to study the  influence of the energy conservation on dynamical critical phenomena \cite{Halperin74}
contains a non-conserved $n$-component order parameter $\vec{\varphi}_0=(\varphi_{0,1},\varphi_{0,2},...,\varphi_{0,n})$  and
a conserved scalar secondary
density $m_0$.  The equations of motion for  $\vec{\varphi}_0$ and  $m_0$ are the following
\begin{eqnarray}\label{eq_mov2}
\frac{\partial {\varphi}_{i,0}}{\partial
t}&=&-\mathring{\Gamma}\frac{\partial {\mathcal H}}{\partial
{\varphi}_{i,0}}+{\theta}_{{\varphi}_{i}}, \qquad i=1\ldots n,\\
\label{eq_mov2v}
 \frac{\partial {m}_0}{\partial
t}&=&\mathring{\lambda}\nabla^2\frac{\partial {\mathcal H}}{\partial
{m}_0}+{\theta_{{m}}} \, .
\end{eqnarray}
The order parameter relaxes and the conserved density diffuses with
kinetic coefficients $\mathring{\Gamma}$, $\mathring{\lambda}$
correspondingly. The stochastic forces ${\theta}_{\varphi_i}$,
${\theta}_{m}$ obey the Einstein relations:
\begin{eqnarray}\label{1}
<\!{\theta}_{\varphi_i}(x,t){\theta}_{\varphi_j}(x',t')\!>\!\!&=&2\mathring{\Gamma}\delta(x-x')\delta(t-t')\delta_{ij},
\\ \label{2}
 <\!{\theta}_{m}(x,t){\theta}_{{m}}(x',t')\!\!>\!&=&\!\!{-}2\mathring{\lambda}\!\nabla^2\!\delta(x{-}x')\delta(t{-}t')
\, ,
\end{eqnarray}
ensuring that static critical properties of the system in $d$ dimensions
are  described by the equilibrium effective static functional
$\mathcal H$:
\begin{eqnarray}
\label{effram} \hspace{-5em} {\cal H}&{=}& \int d^d x
\Bigg\{\frac{1}{ 2} \left[|\nabla
\vec{\varphi}_0|^2{+}\mathring{\tilde r}
|\vec{\varphi}_0|^2\right]\!{+} \frac{\mathring{\tilde u}}{4!}
|\vec{\varphi}_0|^4+\nonumber\\&& \frac{1}{2}{{m^2_0}}+
\frac{1}{2}\mathring{\gamma}
{{m}_0}|\vec{\varphi}_0|^2-\mathring{h}{{m}_0}\Bigg\},
\end{eqnarray}
where $\mathring{\tilde r}$ is connected with relative temperature distance to the critical point, $\mathring{\tilde u}$  and $\mathring{\gamma}$ are static coupling constants and $\mathring{h}$ is an external field.

Integrating out the secondary density one reduces (\ref{effram}) to the
usual Ginzburg-Landau-Wilson model with
new parameters $\mathring{u}$ and  $\mathring{r}$ expressed by the
model parameters $\mathring{\tilde r},\mathring{\tilde u},
\mathring{\gamma}$ and $\mathring{h}$ via the relations:
\begin{equation}
\mathring{r}=\mathring{\tilde r}+\mathring{\gamma} \mathring{h},
\qquad \mathring{u}=\mathring{\tilde u}-3{\mathring{\gamma}^2}.
\end{equation}

A field-theoretical RG description of model  (\ref{eq_mov2})-(\ref{effram}) is  presented in details in Ref. \cite{Folk03}. In the following subsections we only recall its main points.

\subsection{RG functions}
The field-theoretical RG dynamical approach is based on appropriate Lagrangians incorporating equations of motion for the  corresponding dynamical models \cite{Bausch76}. Several  schemes are available for renormalization of those Lagrangians. Here we use results of the minimal subtraction scheme \cite{t'Hooft}, in which the renormalization of model C is well known \cite{Folk03}.

 Within the minimal subtraction scheme, one
introduces renormalization factors $Z_{a}$, $a=\{\{\alpha\},
\{\delta\}\}$, leading to the renormalized parameters
$\{\alpha\}=\{u,\gamma, \Gamma, \lambda \}$ and renormalized
densities $\{\delta\}=\{\varphi,\tilde\varphi,m,\tilde m\}$, where $\tilde\varphi$, $\tilde m$ are auxiliary field densities introduced to obtain the corresponding Lagrangian for the dynamics defined by Eqs. (\ref{eq_mov2})-(\ref{2}).

The behavior of the parameters under renormalization is
described by the flow equations
\begin{equation}\label{fl}
\ell\frac{d\{ \alpha\}}{d \ell}=\beta_{\{\alpha\}} \, ,
\end{equation}
{where $\ell$ is the flow parameter describing the effective critical behaviour.}
The $\beta$-functions for the static  parameters have the
following explicit form:
\begin{eqnarray}
\beta_{u}(u)&=&u f_u(u){=}u(\epsilon+\zeta_{\varphi}(u)+\zeta_{u}(u)),\label{bu}\\
\beta_{\gamma}(u,\gamma)&=&\gamma f_\gamma(u,\gamma){=}\gamma(\frac{\epsilon}{2}{+}\zeta_{\varphi^2}(u){+}\frac{\gamma^2}{2}B_{\varphi^2}(u)),
\label{gam}
\end{eqnarray}
where the function $B_{\varphi^2}(u)$ is obtained from the additive renormalization
$A_{\varphi^2}$ for the specific heat:
\begin{equation}
B_{\varphi^2}(u)=\mu^{\epsilon}Z^2_{\varphi^2}\mu\frac{d}{d
\mu}\left(Z^{-2}_{\varphi^2}\mu^{-\epsilon}A_{\varphi^2}\right) .
\end{equation}
Here, $\mu$  is the scale parameter and factor $Z_{\varphi^2}$
renormalizes  the vertex with $\varphi^2$ insertion.

RG functions $\zeta_{a}$  in (\ref{bu}), (\ref{gam}) describing the critical properties are obtained from $Z$-factors:
\begin{eqnarray}\label{def_z}
\zeta_{a}(\{\alpha\})&=&-\frac{d\ln Z_{a}}{d \ln \mu} \, .
\end{eqnarray}
Relations between the renormalization factors lead to
relations between the corresponding $\zeta$-functions (for details see \cite{Folk03}). In consequence for the
description of the critical dynamics one needs only
$\zeta$-functions of the coupling,
 $\zeta_{u}$,
 the order parameter
$\zeta_{\varphi}$, the auxiliary field $\zeta_{\tilde\varphi}$,
$\varphi^2$-insertion $\zeta_{\varphi^2}$ and also function
$B_{\varphi^2}$. In particular, the $\zeta$-function of the  ratio
\begin{equation}
w=\frac{\Gamma}{\lambda}
\end{equation}
characterizing the time scales of two dynamical densities is related to
the above $\zeta$-functions:
\begin{equation}\label{zWW}
\zeta_{w}(u,\gamma,w){=}\frac{1}{2}\zeta_{\varphi}(u){-}\frac{1}{2}{\zeta_{\tilde\varphi}}(u,\gamma,w){-}\gamma^2
B_{\varphi^2}(u) .
\end{equation}

The dynamical $\beta$-function for the introduced time scale ratio $w$  reads then
\begin{eqnarray}\label{WW}
\hspace{-2em}\beta_{w}(u,\gamma,w)&=&w\,\zeta_{w}(u,\gamma,w)\nonumber\\
&=&w(\frac{1}{2}\zeta_{\varphi}(u){-}\frac{1}{2}{\zeta_{\tilde\varphi}}(u,\gamma,w){-}\gamma^2
B_{\varphi^2}(u)).
\end{eqnarray}

This ratio  $w$ can take values from zero to infinity. Therefore in order
to work in the space of finite parameters  it turns out to be useful to introduce
the parameter $\rho=w/(1+w)$. Then instead of the flow equation   (\ref{fl}) for $w$ the flow
equation for $\rho$ arises:
\begin{equation}\label{drho}
\ell\frac{d\rho}{d\ell}=\beta_{\rho}(u,\gamma,\rho),
\end{equation}
where according to (\ref{WW})
\begin{eqnarray}\label{brho}
\beta_{\rho}(u,\gamma,\rho)&=&\rho(\rho-1)\zeta_{w}(u,\gamma,\rho)\nonumber\\
&=&\rho(\rho-1)\left(\zeta_{\Gamma}(u,\gamma,\rho)-\zeta_{\lambda}(u,\gamma)\right).
\end{eqnarray}
In (\ref{brho})  $\zeta_\Gamma$, $\zeta_{\lambda}$ are the RG functions describing renormalization of the corresponding kinetic coefficients.

The static RG functions of model C are known now in high orders. Within the minimal subtraction scheme five-loop results for functions $\beta_u$, $\zeta_\varphi$, $\zeta_{\varphi^2}$ are accessible \cite{Kleinert91}. The function $B_{\varphi^2}(u)$ is derived within the five-loop order approximation too \cite{Larin98}. While the only dynamical function $\zeta_{\tilde\varphi}$ is known  only within two-loop order \cite{Folk03}. These expressions will serve us below to analyze the phase diagram of model C in the parametric  space of order parameter and space dimensions.

\subsection{Fixed points ant their stability }
The asymptotic critical behavior is analysed  from the
knowledge of the fixed points (FPs) of the flow equations
(\ref{fl}). A FP $\{\alpha^*\}=\{u^*,\gamma^*,\rho^*\}$ is
defined as a simultaneous zero of the $\beta$-functions (\ref{bu}), (\ref{gam}), (\ref{brho}). Eqs. (\ref{bu}) and (\ref{gam}) give static FPs, whereas  Eq.~(\ref{brho}) defines the existence regions of the different possible dynamical FPs.  Checking the structure of (\ref{brho}) one can see that it has three different FPs {(at least within two loop order)}: $\rho^*=0$ , $\rho^*=1$, and $\rho^*=\rho_C$ ($0<\rho_C<1$). The first two solutions exist  for any $n$ and $d$ whereas the solution $\rho^*=\rho_C$ is found from $\zeta_w(u^*,\gamma^*,\rho^*)=0$ and  exists only in a certain region of the  $(n, d)$ plane.
 The stability of these FPs is  defined by (\ref{omrho}) below.
The  FP  which is stable and accessible from the initial conditions corresponds to the specific behavior at
the critical point. Being calculated in this  FP, $\{\alpha^*\}$, the RG functions $\zeta_\Gamma$, $\zeta_\lambda$ define dynamical critical exponents of the order parameter and of the conserved density via relations:
\begin{eqnarray}\label{zzz}
z&=&2+\zeta_{\Gamma}(u^*,\gamma^*,\rho^*),\\
z_m&=&2+\zeta_{\lambda}(u^*,\gamma^*).\label{zmm}
\end{eqnarray}

A FP is stable if all eigenvalues
$\omega_i$ of the stability matrix
${\partial\beta_{\alpha_i}}/{\partial {\alpha_j}}$ calculated at
this FP have  positive real parts. The values of $\omega_i$ indicate
also how fast the renormalized model parameters reach their fixed
point values. From the structure of the  $\beta$-functions (\ref{bu}), (\ref{gam}), (\ref{brho}) we conclude, that the
stability of any FP with respect to the parameters $u$, $\gamma$ and $\rho$
is determined solely by the derivatives of the corresponding
$\beta$-functions:
\begin{eqnarray}\label{omegas}
\omega_{u}(u^*)&=&\left.\frac{\partial\beta_{u}(u)}{\partial
u}\right|_{\{\alpha^*\}},\nonumber\\ \omega_{\gamma}(u^*,\gamma^*)&=&\left.\frac{\partial\beta_{\gamma}(u,\gamma)}{\partial
{\gamma}}\right|_{\{\alpha^*\}},\\ \omega_{\rho}(u^*,\gamma^*,\rho^*)&=&\left.\frac{\partial\beta_{\rho}(u,\gamma,\rho)}{\partial {\rho}}\right|_{\{\alpha^*\}}
\, . \nonumber
\end{eqnarray}

The FP coordinates of model C as well as their stability were  established in Ref. \cite{Folk03}. Here we present FPs which,  being stable for some values of $n$ and $d$, describe  scaling regimes of model C in the corresponding regions of $(n,d)$ plane, see Fig.~\ref{curves} for explicit description.
 FP $\{u_H,0,0\}$ describes the situation when the conserved density is decoupled from the order parameter (region $\mathcal D$), while FP $\{u_H,\gamma_C,0\}$ corresponds to the weak scaling regime, where the order parameter and the conserved density scale with different exponents (region $\mathcal W$).
{  The  strong scaling regime (region $\mathcal S$) is described by the FP $\{u_H,\gamma_C,\rho_C{<}1\}$, which exists only at certain values $n$ and $d$, as was noted already. In this regime both quantities have the same critical exponent $z=z_m$, that is the consequence of $\zeta_w(u_H,\gamma_C,\rho_C{<}1)=0$, therefore at this FP the $\zeta$-functions for the kinetic coefficients are equal to each other (see Eqs. (\ref{brho}) and (\ref{zzz}),(\ref{zmm})).}
   { Region $\mathcal A$, where the behavior of the secondary density is much slower than the behavior of the order parameter, corresponds to the stability region of the FP $ \{u_H, \gamma_C, 1\}$. However it appears to be unstable within perturbative field-theoretical RG \cite{Folk03}, therefore  region  $\mathcal A$ does not exist within the perturbative field-theoretical RG.}

It is known (see e.g. \cite{Folk03}) that $\omega_\gamma$ governs the stability of FPs $\{u_H,0,0\}$ and  $\{u_H,\gamma_C,0\}$. Depending on $n$ and $d$ the stability exponent  $\omega_\gamma$  will be positive for $\{u_H,0,0\}$ and negative for $\{u_H,\gamma_C,0\}$ or {\em vice versa}. Therefore the condition of vanishing $\omega_\gamma$  defines a border between stability regions for FP $\{u_H,0,0\}$ and  FP $\{u_H,\gamma_C,0\}$:
\begin{equation}\label{cond0}
\omega_\gamma(u_H,0)=0.
\end{equation}
It can be shown \cite{Folk03} that this condition is equivalent to  the vanishing of the specific heat exponent $\alpha$:
\begin{equation}\label{cond1}
\alpha(n,\epsilon)=0.
\end{equation}
Therefore, this condition gives a border line $n_\alpha(\epsilon)$ (or $\epsilon_\alpha(n)$) in the parametric space $(n,d)$ between
the regions  $\mathcal D$ and $\mathcal W$.

The border between the stability of  the FPs $\{u_H,\gamma_C,0\}$ and  $\{u_H,\gamma_C,\rho_C\}$ is governed by the stability exponent with respect to $\rho$:
\begin{eqnarray}\label{omrho}
\omega_\rho(u^*,\gamma^*,\rho^*)&=&(1-2\rho^*)\zeta_w(u^*,\gamma^*,\rho^*)+\nonumber\\&&\rho^*(1-\rho^*)\left.\frac{\partial\zeta_w(u,\gamma,\rho)}{\partial \rho}\right|_{\{\alpha^*\}}.
\end{eqnarray}
{ Note, that for $\rho^*=1$ the transient exponent diverges as $\ln(1-\rho^*)$. Thus $\rho^*=1$ is nowhere stable at least in two loop approximation (see Eq. (85) in the first reference of \cite{Folk03}).  If $\zeta_w(u^*,\gamma^*,\rho^*)$ is zero the FP is marginal.}
Therefore, values of $n$ and $d$ at which the condition
\begin{equation}\label{omrho00}
\omega_\rho(u_H,\gamma_C,0)=0
\end{equation}
is satisfied give us the border between stability regions of the FPs $\{u_H,\gamma_C,0\}$ and  $\{u_H,\gamma_C,\rho_C\}$. It was shown \cite{Folk03} that this condition is equivalent  to  the condition
\begin{equation}\label{cond2}
c\eta=\frac{\alpha}{\nu},
\end{equation}
in all orders of perturbation theory, where $c\eta=z-2$ is a part of the dynamical critical exponent of model A.
Thus (\ref{cond2}) defines a border $n_1(\epsilon)$  (or  $\epsilon_1(n)$) between  the weak scaling region $\mathcal W$ described by the FP $\{u_H,\gamma_C,0\}$ and the strong scaling region $\mathcal S$  described by the FP $\{u_H,\gamma_C,\rho_C\}$.

{Conditions (\ref{cond1}) and (\ref{cond2})  are valid in all orders of perturbation theory and they include  critical exponents of $O(n)$ symmetrical model. Critical exponents are universal quantities depending only on the global characteristics as space dimension $d$ and order parameter dimension $n$. Within  RG critical exponents can be calculated as functions of $d$ and $n$. Therefore studying the dependence of critical exponents on $d$ and $n$ we can extract from (\ref{cond1}) and (\ref{cond2}) the border lines between regions $\mathcal D$, $\mathcal W$, $\mathcal S$ in $(n,d)$ plane. Whereas condition (\ref{cond1}) is purely static and separates the region, where the static coupling $\gamma$ is relevant or it is not, condition (\ref{cond2}) is a dynamical condition. It separates in the region where the static coupling $\gamma$ is relevant and
the order parameter follows the model A dynamics from the region where the  genuine model C dynamics is present. 
It is therefore possible to improve the two-loop result for these two borderlines by using higher order field theoretical results for the $\phi^4$ theory and model A alone.
}

 There are two alternative ways to analyse perturbative  RG functions in order to get universal quantities. Within the first approach one applies $\epsilon$-expansion to obtain the corresponding quantities in a form of series in $\epsilon$ and then to evaluate them at the value of interest. Within the second way of analysis one fixes the space dimension $d$ to a certain value and then directly solves the  system of equations for FPs  numerically, the so-called {\em fixed dimension approach}. In the next two sections we use these approaches to analyze conditions (\ref{cond1}) and (\ref{cond2}) numerically.

\section{High order $\epsilon$-expansions for border lines \label{III}}

Conditions (\ref{cond1}) and (\ref{cond2}) include  static critical exponents as well as the dynamical critical exponent of model A for the $O(n)$ symmetrical model. These quantities are known now within high orders of perturbative field-theoretical RG.  Therefore we can use these expressions  to study conditions (\ref{cond1}) and (\ref{cond2}).
In particular, the condition  determining $\epsilon_\alpha(n)$ (\ref{cond1}) coincides with the condition for  the marginal dimension $n_c$ of
a weakly diluted $O(n)$ model according to the Harris criterion \cite{Harris74,Holovatch01}.
We can use the known $\epsilon$-expansion for  $n_c\equiv n_\alpha$ obtained on the base of five-loop  minimal subtraction RG   functions \cite{Holovatch01}:
\begin{equation}\label{nalpha}
n_{\alpha}=4 - 4 \epsilon + 4.707199 \epsilon^2 - 8.727517 \epsilon^3 +
 20.878373 \epsilon^4.
 \end{equation}

Now we consider condition (\ref{cond2}). Note that model A quantity,   $c \eta$,  for a system with $O(n)$ symmetrical order parameter  is now known in four-loop order within the minimal subtraction RG scheme \cite{Adzhemyan08}, while expressions for static  critical exponents for this model  are obtained in the next fifth order\cite{Kleinert91}. To be consistent, we restrict ourselves only to the four-loop  expressions for the static exponents. Substituting them into (\ref{cond2}) and then  keeping the coefficients
as functions of $n$ and reexpanding the (\ref{cond2}) in $\epsilon$ we obtain $n_1$ in the form:
\begin{equation}\label{n1}
n_1=4 - 4.181523 \epsilon + 4.751724 \epsilon^2 -
 8.701434 \epsilon^3.
 \end{equation}

Formally, numerical values for $n_\alpha$ (\ref{nalpha}) and $n_1$ (\ref{n1}) at given space dimension  may be obtained by fixing the value of $\epsilon$. However, the series for RG functions are known to be of asymptotic nature \cite{asymptotic1,asymptotic2} therefore some resummation procedure should be applied to extract reliable information on their basis. Different resummation procedures were successfully applied for numerical analysis of $\epsilon$-expansions  obtained for some marginal dimensions (stability borders) of static field-theoretical models \cite{Holovatch01,Dudka04,Dudka12}.
To get border lines from (\ref{nalpha}) and  (\ref{n1}) we can apply to them a Pad\'e-Borel resummation technique.
 Such  resummation procedure  has been successfully used
in various tasks of theory of critical phenomena \cite{Pade}. The procedure
is based  on the integral Borel transformation
\cite{Hardy48}, and
uses  an extrapolation by means of a Pad\'e-approximant as an intermediate step
\cite{Baker81}.
 Starting from the initial sum $S$ of $L$ terms $S(x)=\sum_{i=0}^La_i x^i$ we construct its
Borel image
\begin{equation}
\label{BLimage}
 S^{\rm B}(xt)=\sum_{i=0}^L\frac{a_i(x t)^i}{i!}.
\end{equation}
Subsequently we extrapolate the Borel image
(\ref{BLimage}) by a rational
Pad\'e approximant:
\begin{equation}\label{Pad}
S^{\rm B}(xt)\quad\Rightarrow\quad  \left[ L-1/1 \right](xt)=\frac{\sum_{i=0}^{L-1} b_i (xt)^i}{(1+c_1 xt)},
 \end{equation}
 with the coefficients $b_i, c_1$ expressed in terms of the initial coefficients $a_i$ and  the denominator linear in $xt$ to avoid problems with multiple poles.

The resummed function $S^{\rm Res}$ is finally  obtained by the inverse Borel transform:
\begin{equation}
\label{res}
S^{\rm Res}(x)=\int_0^\infty\, dt \exp (-t)
\left[ L-1/1 \right] (x t).
\end{equation}

Applying (\ref{BLimage})-(\ref{res}) with $x=\epsilon$ to   (\ref{nalpha}) and then fixing the value of $\epsilon$  we can obtain the corresponding  numerical value $n_\alpha$, while doing the same   for (\ref{n1}) we can get the numerical value of $n_1$.
However note, that expansion (\ref{n1}) is one order shorter than (\ref{nalpha}). To obtain border lines within the same order we neglect $\epsilon^4$ term for $n_\alpha$ (\ref{nalpha}).
Therefore applying the Pad\'e-Borel procedure (\ref{BLimage})-(\ref{res}) to (\ref{nalpha}) without the last term  and to (\ref{n1}) and changing $\epsilon$ from 0 to 2 we get  border lines  $n_\alpha(\epsilon)$ and  $n_1(\epsilon)$ given by the green solid and dashed curves respectively  in   Fig.~\ref{curves}a.
As one can see, the region for the weak scaling regime is larger, compared to the two-loop order results \cite{Folk03}, and in qualitative accordance with  the NPRG results \cite{Mesterhazy13}. Such border lines  are very close to the nonperturbative results in the vicinity of $\epsilon=1$.

\section{Resummation of the RG functions at fixed space dimension\label{IV}}
\subsection{Borderlines}

As it was already noted, static RG functions are known up to the five loop order  within the minimal subtraction  RG scheme \cite{Kleinert91}, as well as the    factorial divergence of their coefficients in coupling $u$ is established \cite{asymptotic1}.
Therefore we apply  the procedure (\ref{BLimage})-(\ref{res}) with $x$  meaning $u$ now
 to the five-loop static RG function $f_u(u)$, as well as to the polynomials $f_\gamma(\gamma,u)-\gamma^2 B_{\varphi^2}(u)/2=\frac{\epsilon}{2}+\zeta_{\varphi^2}(u)$ and  $\zeta_\varphi(u)/u^2$. Note that in our calculation we use the two-loop expression for $B_{\varphi^2}(u)$: $B_{\varphi^2}(u)=n/2$.

Although we work in {\em fixed dimension approach}, in order to get the corresponding border lines we fix the value of $n$ and then solve a system of equations. In particular, solving the system of equations $f_u(u_H)=0$ and (\ref{cond0}), where the corresponding functions are substituted by their resummed counterparts for fixed values of $n$ we find the stability border line $\epsilon_\alpha(n)$ separating in the $(n,\epsilon)$ plane the decoupled scaling region $\mathcal D$ from weak scaling region $\mathcal W$. The result is shown in Fig.~\ref{curves}b by a blue solid curve.

In a similar way, solving the system of equations $f_u(u_H)=0$, $f_\gamma(u_H,\gamma)=0$ and (\ref{omrho00}), containing resummed functions for fixed values of $n$ we find the stability border $\epsilon_1(n)$ separating in the $(n,\epsilon)$ plane the weak scaling region $\mathcal W$ from the strong scaling region $\mathcal S$. The result is given in Fig.~\ref{curves}b by a  blue dashed curve.

The agreement of these results with those of the NPRG approach \cite{Mesterhazy13} (black curves in Fig.~\ref{curves}) is worse than for  the   $\epsilon$-expansion for the border lines (green curves in Fig.~\ref{curves}a). Nevertheless the  qualitative shift up in values  of $\epsilon$ for fixed $n$ is present in the region $n>1$ and $\epsilon<1$ compared to the two-loop order result. Also   the region  $\mathcal W$ for the weak scaling regime is wider then  within the two loop approximation \cite{Folk03}. A possible existence of an anomalous regime $\mathcal A$ is considered in next subsection.

\subsection{Stability of the FP describing $\mathcal S $ regime and possible existence of region $\mathcal A$ }

The existence of the region $\mathcal A$ in $(n,d)$ plane  is connected with the behavior of the FP $\{u_H,\gamma_C,\rho_C\}$. For some values of $n$ at  small $\epsilon$ the value of $\rho_C$ has the tendency to go to unity,  approaching the FP $\{u_H,\gamma_C,1\}$, which was shown to be unstable within the two-loop approximation \cite{Folk03}.  Also it was proven that in this loop order  the value of $\rho_C$ being very close to 1 never reaches this value.
Nevertheless, note that $\rho\to 1$ means the time scale ratio $w \to \infty$, therefore the behavior of the order parameter is much faster than that of the conserved density.
However the dynamical critical exponents for  the order parameter and for the conserved density remain the same for all values of  FP $\{u_H,\gamma_C,\rho_C\}$.

\begin{figure}
\includegraphics{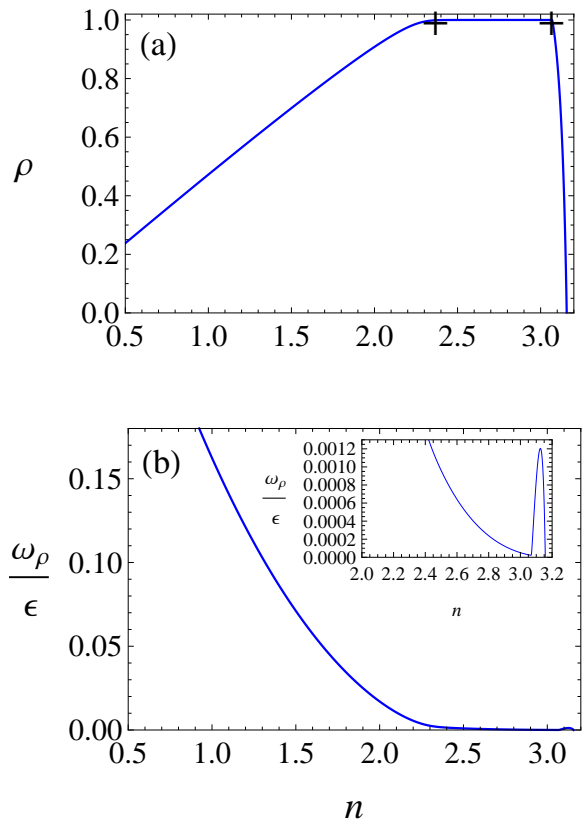}
\caption{ Values of $\rho_C$ calculated from $\zeta(u_H,\gamma_C,\rho_C)=0$ (a) as well as $\omega_\rho(u_H,\gamma_C,\rho_C)$ calculated from (\protect\ref{omrho}) (b) at $\epsilon=0.25$ as functions of the order parameter  dimension $n$ obtained on the base of the resummed five-loop static RG functions. The inset depicts the right  bottom part of the corresponding picture. { The marks '+' denote artificial region for $\epsilon=0.25$ shown by black dots in Fig.~\protect\ref{curves}}
}
\label{rho_omega}
\end{figure}

We check  these results of Ref.  \cite{Folk03} looking for FP solutions  of the beta functions  (\ref{bu}), (\ref{gam}), (\ref{brho}) with all non-zero coordinates  $\{u^*,\gamma^*,\rho^*\}$ using the same  resummed five-loop functions as in the subsection above. We calculate also the critical exponent $\omega_\rho$ for this FP. Our results for $\rho_C$ and $\omega_\rho$ obtained at fixed $\epsilon=0.25$ as functions of $n$ are shown in Fig.~\ref{rho_omega}.

Fig.~\ref{rho_omega} qualitatively supports the situation observed already within the two-loop order \cite{Folk03}. The value of $\rho_C$ numerically is always less than 1, however one does not see this within a resolution of Fig.~\ref{rho_omega},  while $\omega_\rho$ is always positive. That means stability of the FP $\{u_H,\gamma_C,\rho_C\}$ in the full region of existence.  These results were expected, because only two loop expressions for dynamical RG functions are accessible: Taking into account   high-loop orders for static functions  only does not  lead to considerable changes of the results for the dynamical  values.

However looking at our solution for the time scale ratio in the region $\mathcal A$ found in the NPRG approach we observe the following.
Increasing $n$ from the small values to
larger ones at fixed dimension up to the boundary  where the decoupling region is reached, $\rho_C$ increases first and then decreases showing a maximal value.
This maximum is always smaller than one, however below  $\epsilon=0.4$  this maximum is of order $\rho_C\approx 0.99$. Decreasing $\epsilon$ further $\rho$ seems to develop a plateau value at almost 1 over a certain region of $n$ (see Fig.~\ref{rho_omega}), which is qualitatively in agreement with the existence  of a region $\mathcal A$.

To be specific we define a region where $\rho_C$ rises to the value of almost one. Let us take the
values $n$ when $\rho_C$ is crossing 0.999 as a left part of a border for
this artificial region. Then we  choose then position of  the 'beak' shown in
the inset of Fig.~\ref{rho_omega} as the right part of the border. Data obtained in this way are shown in Fig.~\ref{curves}  by black dots   for
values $\epsilon$ from 0.05 till 0.35 with a step size of 0.05.  For $\epsilon=0$ we plot one loop values  $n=2$ and $n=4$. This region is also marked by '+' in the upper picture of Fig.~\ref{rho_omega}. As one can see in Fig.~\ref{curves} this artificial region turns out to be roughly similar to region ${\mathcal A}$ found in the NPRG analysis\cite{Mesterhazy13}, however  increasing the limiting value for border from $\rho=0.999$ to larger values below 1 one shifts the left border
line to larger values of $n$. In the limit $\rho \to 1$ this region of course disappears in our approach.


\section{Discussion and conclusions \label{V}}
In this study we have reexamined perturbative field-theoretical results obtained for model C within two-loop order without resummation.
In order to do this we have used already  known five-loop RG expressions for the field-theoretical $O(n)$ model, as well as $\epsilon$-expansions  known within high-loop orders for marginal dimension of the diluted $O(n)$ model and for the  dynamical critical exponent of model A.
In particular, using four-loop $\epsilon$-expansion for $c\eta$ we have derived the $\epsilon$-expansion for border line  $n_1$, that separates weak and strong scaling regimes of model C  up to $O(\epsilon^4)$.   Our result is compared with the result of NPRG approach \cite{Mesterhazy13} obtained recently for the model C dynamics.

Our analysis qualitatively supports NPRG results \cite{Mesterhazy13} for $\epsilon \le 1$ and $n\ge 1$, leading to the wider region for weak scaling regime and larger values $\epsilon_\alpha(n)$ and  $\epsilon_1(n)$ in comparison with results obtained on the base of two-loop RG functions without resummation. However there are striking deviations in other regions of the $(n,\epsilon)$-plane. Note that $\varphi^4$ models are not appropriate for investigations for $d<3$, because of a different physics in low dimensions. Moreover,  for $\epsilon>1$ (that is $d<3$) it is obvious that $\epsilon$-expansion does not work because the expansion parameter $\epsilon$ now  is not small.  While in the NPRG study \cite{Mesterhazy13} only the local potential approximation with scale dependent constant at gradient term (LPA') was used with the truncation of a local potential of $\varphi^4$ type. This approximation relies on the assumption that correlation functions with large number of legs  have a small impact on the RG flow as those with fewer legs, as well as on the assumption that the anomalous dimension is small. This is not  the case for the low-dimensional systems ($d<3$), therefore high-order field expansions should be used to go below $d=3$.

Finite truncations within the NPRG approach have another unpleasant effect: they induce a residual dependence of the physical quantities on the choice of a cut-off function used to suppress  low-momentum fluctuations.  The NPRG approach of Ref.\cite{Mesterhazy13}  uses a sharp $\theta$-cut-off, which, although  nonanalytical in its nature, leads to analytical expressions of non-perturbative $\beta$-functions. However other cut-off functions like a power-law or an exponential  can be used. For instance, data of a NPRG study of the two-dimensional $O(2)$ model  with the help of an optimized exponential cut-off function are in very good agreement with universal features of the Kosterlitz-Touless transition \cite{Jakubczyk14}. Probably the non-monotonic behavior of the border line between regions $\mathcal W$ and $\mathcal S$  obtained in Ref.  \cite{Mesterhazy13} (black dashed curve in Fig.~\ref{curves}) is a consequence  of  the use of the $\theta$-cut-off function.  In any case it is interesting to have data from other choice of  a cut-off function. The check of model C behavior at $n\to 0$ proposed in  Ref. \cite{Mesterhazy13}  on the base of Monte Carlo simulations for  SAW models in fractal dimensions is not reliable, since the similarity between SAW and $O(n\to 0)$ model was proven only for the static case \cite{deGennes}. Moreover, the relation between the non-integer dimension arising due to the analytic continuation in field theories (e.g. via $\epsilon$-expansion) and fractal dimension is not straightforward \cite{fractal_dim}.

Conditions that govern the borders between regions $\mathcal D$ and $\mathcal W$ (\ref{cond1}) as well as between  $\mathcal W$ and $\mathcal S$  (\ref{cond2}) are valid to all orders of perturbation theory. Therefore it can be also checked within the NPRG approach on the base of a simpler  $O(n)$-model and dynamical model A. However model A critical dynamics was studied within NPRG  for a scalar order parameter only\cite{Canet}. Moreover, despite recent NPRG studies of $O(n)$ models in fractional dimensions \cite{Codello13,Mati15}  the condition for a vanishing  exponent $\alpha$ was not  a subject of  interest.

The most intriguing point and a genuine model C feature is  that NPRG results are in favor of the existence of an anomalous region $\mathcal A$. Within this approach  region $\mathcal A$ is described by a stable solution with $\gamma\not = 0$ and $\rho=1$, as well as with a critical exponents $z<z_m$. This result requires its confirmation by studies with more elaborate truncations as well as with other choices of a cut-off function. Concerning the  perturbative field-theoretical RG approach, region $\mathcal A$ can be checked only if high-loop order dynamical RG functions will be accessible. They can possibly  lead to new solutions for dynamical FPs or a change of   stability of the present found FPs. However one should be careful trusting new solutions obtained only in high orders of perturbations, because they might be   controversial (see {\em e.g.} Ref.\cite{frustrated}, {where one can find a discussion about conflicting results of perturbed field-theoretical RG and NPRG approaches for frustrated magnets}).

This work was supported in part by the 7th FP, IRSES Projects No. 295302
`Statistical Physics in Diverse Realizations', No. 612707 `Dynamics of and in
Complex Systems', No. 612669 `Structure and Evolution of Complex Systems with
Applications in Physics and Life Sciences'. M.D. thanks to Danylo Dobushovskii
for advice with some technical points.

\end{document}